# A direct test for instantaneous collapse of wave functions in configuration space

Avi Marchewka

Galei Techelet 8, Herzliyya, Israel.

## Abstract

Wavefunction collapse is a puzzling aspect of quantum mechanics. Designing a test to confirm the instantaneousness of collapse has turned out to be challenging, especially for collapse in configuration space. We propose a test using two identical, non-local, correlated photons in an interferometer in which a post-selection measurement of one of the photons in its location changes the statistical behavior of the other photon, which is an arbitrarily large distance away, and thus its detectable subsequent behavior. Analysis of the resulting correlations constitutes a test of the instantaneousness (or non-instantaneousness) of collapse. Connections to some of the recent proposed models of collapse are discussed.

## 1. Introduction

Collapse of the wavefunction due to measurement is a central assumption in quantum mechanics [1]. However, it is far from fully understood or even agreed on [2]. What goes on in the measurement process that brings about the collapse of the wavefunction, and what makes it instantaneous, if it is indeed so? Part of the difficulty with collapse is that any physical process takes a finite amount of time, while collapse is conceived of as instantaneous, as the wavefunction jumps from one state to another. There have been many approaches to quantum theory that either evaded the use of the collapse, *e.g.*, Ref. [3], or modified it, *e.g.*, Ref. [4]. Furthermore, collapse of the wavefunction in configuration space has unphysical behavior derivative of the instantaneous collapse of the wavefunction, mainly its inferred superluminous propagation and divergence of the particles' average displacements, which is inconsistent with the static interpretation of the wave function [5]. The assumption of collapse raises the fundamental question of whether a wave function should be regarded as an epistemic or an ontological entity [3,6]. Hence, a direct and explicit test for the instantaneousness or finite duration of collapse of a wavefunction in configuration space is important.

The theoretical study and experimental tests for the instantaneous collapse duration has a long history [7], dating at least from Einstein's suggestion at the 5th Solvay Conference in 1927 [8]. This suggestion of Einstein led to photon anti-bunching experiments [9,10]. In these experiments, a single-photon wavefunction is split into two, each outgoing wavefunction going into one arm of a beam splitter. Then, correlation detections of the photon or one of its properties is used to test the collapse and to establish limits on the duration of collapse. However, there is not yet a *direct* measurement that shows the instantaneousness of collapse. This kind of test is given here. The working principal of the gedanken experiment herein may be simplified as follows. Consider two entangled identical photons; post-selected measurement on one of the photons collapses the second photon into a different state. This difference is used to estimate the collapse duration.

In Section 2, we will show the physical setup of the interferometer, and describe the experimental procedure. In Section 3, to calculate the different correlation functions between the photon detections in the two detectors, three distinct cases will be considered. In Section 4, we will show how to test the instantaneousness of the collapse in configuration space. In Section 5, a few of the ideas suggested in Ref. [6] are discussed in view of the gedanken experiment herein. Section 6 has a summary and discussion.

## 2. Interferometer setup

Figure 1 shows the interferometer setup that will be used to test instantaneousness of the collapse. The capital letters *A*, *B*, *C*, and *D* without subscripts, denote four symmetrical beam splitters. Lower-case letters $a_1$, $a_2$, $b_1$, $b_2$, $c_1$, $c_2$, $d_1$, $d_2$ denote particle paths. Capital letters $D_j$ with subscripts $j=\{c_1, c_2, d_1, d_2\}$ label the detectors; the subscripts correspond to the particle paths at the end of which the detectors are placed. The lengths between the beam splitters are all equal to one another,

$$a_1 = a_2 = b_1 = b_2 \equiv L. \tag{1}$$

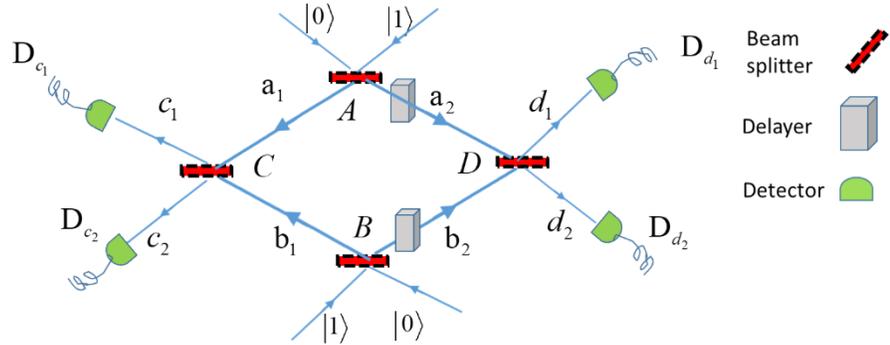

The distances from the beam splitters $C$, $D$ to their adjacent detectors $D_j$, $j=\{c_1, c_2, d_1, d_2\}$ are also all equal to one another,

$$c_1 = c_2 = d_1 = d_2 \equiv l. \tag{2}$$

Two equal time delay components, denoted by $\tau$, are set at arm $\overline{AD}$ ($a_2$) and at arm $\overline{BD}$ ($b_2$). The delay time can take any non-negative value, $\tau \geq 0$.

The kets $|\#\rangle_i$ denote the number of photons, $\# = \{0,1\}$. The subscripts $i = \{a_1, a_2, b_1, b_2, c_1, c_2, d_1, d_2\}$, denote the 8 possible legs that the photons may be in. The commutation relations for photons are

$$[\hat{a}_j, \hat{a}_k^\dagger] = \delta_{i,k}$$
$$[\hat{a}_j, \hat{a}_k] = [\hat{a}_j^\dagger, \hat{a}_k^\dagger] = 0 \qquad (3)$$

The number operator is $\hat{N}_i = \hat{a}_i^\dagger \hat{a}_i$,

$$\hat{N}_i |n\rangle = n_i |n\rangle \qquad (4)$$

The general matrix for a beam splitter is

$$\hat{U}_{bs} = \begin{pmatrix} t' & r \\ r' & t \end{pmatrix}, \qquad (5)$$

with $|r'| = |r|$, $|t'| = |t|$, $|r|^2 + |t|^2 = 1$, $r^* t + r' t'^* = 0$, and $r^* t + r' t'^* = 0$. For a symmetric beam splitter, $r_k = i/\sqrt{2}$ and $t_k = 1/\sqrt{2}$. For the interferometer in Fig. 1, all beam splitters are taken to be symmetrical.

For clarity, let us use two simplifications in treating the concept of the photon:

- Photons are assumed to propagate at the speed of light. After a photon passes a beam splitter, say $A$, then after time interval $\Delta t$, its position is given by

$$\Delta x = \Delta t \cdot c,$$

where $c$ is the speed of light. For example, it follows that the time a photon spends in an interferometer arm, say $\overline{AC}$ ($a_1$), is $L/c$.

- We assume that we may approximate a photon as a point entity. A more fully correct treatment requires a wavepacket description. Using wavepackets for photons adds a statistical consideration, but this will not change the essential conclusions.

Consider propagation of two indistinguishable photons in the interferometer in Fig. 1. This description will be in the lab frame, according to a universal clock.

Two photons enter the interferometer simultaneously, one initially to the beam splitter A and the other initially to the beam splitter B. Photons that travel from the beam splitters $A$ and $B$ to either of the detectors $D_{c_1/c_2}$ have a traveling time of $(L+l)/c$; photons that travel to detectors $D_{d_1/d_2}$ have travel times that include the delays in legs $a_2$ and $b_2$, that is, $(L+l)/c + \tau$.

The results we are interested in are the post-selected cases in which the first photon is detected on the left side of the interferometer, *i.e.*, in one of the detectors $D_{c_1}$ or $D_{c_2}$, without traversing a delay component; and a second photon is detected on the right side of the interferometer, *i.e.*, in one of the detectors $D_{d_1}$ or $D_{d_2}$, after traversing a delay component. For our purposes, it is sufficient to consider the post-selected case where one photon is detected by $D_{c_1}$ (without a delay) and a second photon is detected at either by $D_{d_1}$ or $D_{d_2}$ (after a delay). Other detection results need not be considered here.

### 3. Three experimental scenarios

Consider running the experiment -- propagating the photons via the interferometers -- with three different values of a time delay $\tau$.

I. No time delay: $\tau_1 = 0$. The photons arrive at the detectors simultaneously, that is, results of interest are either the $D_{c_1}, D_{d_1}$ detections or the $D_{c_1}, D_{d_2}$ detections.

II. Intermediate time delay: $0 < \tau_2 < l/c$. A photon that ends up in detector $D_{c_1}$ will be detected $\tau_2$ earlier by either $D_{d_1}$ or $D_{d_2}$.

III. Long time delay: $\tau_3 > l/c$. A photon that ends up at detector $D_{c_1}$ will be detected earlier by $\tau_3$ from the photon detectors at either detectors $D_{d_1}$ or at $D_{d_2}$.

Note the similarities between Scenario I and Scenario II, and the differences as compared to Scenario III. In the first two scenarios, when a photon is registered at one of the detectors on the left side of the interferometer, the second photon *has already passed* the beam splitter $D$. In contrast, in Scenario III, when a photon registers at one of the detectors on the left side of the interferometer, the second photon *has not yet passed* the beam splitter $D$. Note that the size of the interferometer can be arbitrarily large. The detection can take place long after the two photons have passed beam splitters $A$ and $B$. In principle, there is no barrier to the distance being, for example, one light year. We will see that this difference causes a change in the correlation of the two photons' detection. This will be used below to indicate the instantaneousness of the collapse.

### 3.1 Scenario I: No delay, $\tau_1 = 0$

Let us calculate the wave function of the two photons in the interferometer. The routes for the indistinguishable photons entering at the symmetric beam splitter $A$ are

$$A: |1\rangle_A \to \frac{1}{\sqrt{2}} (|1\rangle_{a_1} + i|1\rangle_{a_2})$$

$$\xrightarrow[D]{C} \frac{1}{2} (i|1\rangle_{c_1} + |1\rangle_{c_2} - |1\rangle_{d_1} + i|1\rangle_{d_2}) \quad (6)$$

and the normalized wavefunction for the photon that started at beam splitter $B$ is

$$B: |1\rangle_B \to \frac{1}{\sqrt{2}} (i|1\rangle_{b_1} + |1\rangle_{b_2})$$

$$\xrightarrow[D]{C} \frac{1}{2} (i|1\rangle_{c_1} - |1\rangle_{c_2} + |1\rangle_{d_1} + i|1\rangle_{d_2}) \quad (7)$$

Then, the wavefunction of the joined photons is given by the tensor product

$$|\psi\rangle = \frac{1}{4\sqrt{N}} (i|1\rangle_{c_1} + |1\rangle_{c_2} - |1\rangle_{d_1} + i|1\rangle_{d_2}) \otimes (i|1\rangle_{c_1} - |1\rangle_{c_2} + |1\rangle_{d_1} + i|1\rangle_{d_2}) \quad (8)$$

Where the factor $\sqrt{N}$ is the normalization of the joint wave function. Eq. (8) can be reduced to the following

$$|\psi\rangle = \frac{1}{4\sqrt{N}} (-|1\rangle_{c_1}|1\rangle_{c_1} - |1\rangle_{c_2}|1\rangle_{c_2} - |1\rangle_{d_1}|1\rangle_{d_1} - |1\rangle_{d_2}|1\rangle_{d_2}$$

$$- 2|1\rangle_{c_1}|1\rangle_{d_2} + 2|1\rangle_{c_2}|1\rangle_{d_1}) \quad (9)$$

In terms of creation operators, the joint wavefunction is

$$|\psi\rangle = \frac{1}{4\sqrt{N}} (-(\hat{a}_{c_1}^\dagger)^2 - (\hat{a}_{c_2}^\dagger)^2 - (\hat{a}_{d_1}^\dagger)^2 - (\hat{a}_{d_2}^\dagger)^2 - 2\hat{a}_{c_1}^\dagger \hat{a}_{d_2}^\dagger + 2\hat{a}_{c_2}^\dagger \hat{a}_{d_1}^\dagger) |0\rangle \quad (10)$$

where by normalization $|\langle\psi|\psi\rangle|^2 = 1$, and $N = 3/4$.

There are few points to be noted in regard to Eqs. (9) and (10).

- The wavefunctions (9) and (10) may be regarded as discrete versions of the "bunching" and "anti-bunching" phenomena that appear in the continuous case in Refs. [11] and [12].
- The behavior of states $|1\rangle_{c_2} |1\rangle_{d_1}$ and $|1\rangle_{c_1} |1\rangle_{d_2}$ are the focus here. According to Eq. (10), if a single photon is detected in, say, $D_{c_1}$, then the other photon will be detected at, correspondingly, $D_{d_1}$, and *vice versa*.

- The lack of states $|1\rangle_{c_2} |1\rangle_{d_2}$ and $|1\rangle_{c_2} |1\rangle_{d_2}$ means that no photon can be detected in those states. This seems surprising, purely the effect of the interference between the two photons.

From Eqs. (9), (10), and (4) we have the following detection correlations

$$\langle\psi|\hat{N}_{c_1}\hat{N}_{c_2}|\psi\rangle = \langle\psi|\hat{N}_{c_1}\hat{N}_{d_1}|\psi\rangle = 0$$
$$\langle\psi|\hat{N}_{d_1}\hat{N}_{d_2}|\psi\rangle = \langle\psi|\hat{N}_{d_2}\hat{N}_{c_2}|\psi\rangle = 0 \tag{11}$$

and

$$\langle\psi|\hat{N}_{c_1}\hat{N}_{d_2}|\psi\rangle = \langle\psi|\hat{N}_{c_2}\hat{N}_{d_1}|\psi\rangle = \frac{1}{3}. \tag{12}$$

Let us define a function to describe the measurement results in detector $j$

$$D_j = \begin{cases} 1 & \text{positive detection} \\ 0 & \text{negative detection} \end{cases}, \tag{13}$$

and define the correlation function between two detectors

$$\text{corr}(D_j, D_{j'}) = D_j \cdot D_{j'} \quad \{j, j'\} \in \{c_1, c_2, d_1, d_2\}. \tag{14}$$

Then, also using Eqs. (11) and (12), the following correlation functions between the detectors hold:

$$\text{corr}_{\tau_1=0}(D_{c_1}, D_{d_1}) = \text{corr}_{\tau_1=0}(D_{c_2}, D_{d_2}) = 0, \tag{15}$$

$$\text{corr}_{\tau_1=0}(D_{c_1}, D_{c_2}) = \text{corr}_{\tau_1=0}(D_{d_1}, D_{d_2}) = 0, \tag{16}$$

and

$$\text{corr}_{\tau_1=0}(D_{c_1}, D_{d_2}) = \text{corr}_{\tau_1=0}(D_{c_2}, D_{d_1}) = 1. \tag{17}$$

For the next steps, any of the correlation functions (15), (16), or (17) could be used. We arbitrary chose to make use of the correlations $\text{corr}(D_{c_1}, D_{d_2})$ and $\text{corr}(D_{c_1}, D_{d_1})$ and their dependence on the time delay.

## 3.2 Scenario II: time delay $0 < \tau_2 < \dfrac{l}{c}$

The delays at arms $a_2$ and $b_2$ have the value $\tau_2$. Consider, while running the experiment, the case where a single photon is detected on the *left* side of the interferometer by detector $D_{c_1}$. In this case, according to Eq. (10), the second photon must be on the *right* side of the interferometer. At the time of the detection of the first photon on the left side, the second photon must have already passed the beam splitter $D$ but not yet have reached either of the detectors $D_{d_1}$ or $D_{d_2}$. Therefore, the second photon wavefunction is deduced by collapsing the joint photon wave function.

That is, using Eq. (10), we have

$$|\psi\rangle \xrightarrow[D_{c_1}=1]{\text{single detection}} |\psi'\rangle = \hat{a}_{d_2}^\dagger |0\rangle \qquad (18)$$

and

$$\langle \psi' | \hat{N}_{d_2} | \psi' \rangle = 1.$$

That is, the second photon is eventually detected at the detector $D_{d_2}$. It follows that its correlation function is the same as in the case of no delay time,

$$\begin{aligned} \text{corr}_{\tau_2}(D_{c_1}, D_{d_2}) &= 1 \\ \text{corr}_{\tau_2}(D_{c_1}, D_{d_1}) &= 0 \end{aligned} \qquad (19)$$

Thus, the correlation function between the detectors $D_{c_1}$ to $D_{d_2}$ and between $D_{c_1}$ and $D_{d_1}$ are the same in Scenario I, where there is zero time delay, and in Scenario II, where the time delay is $\tau_2$.

## 3.3 Scenario III: time delay $\tau_3 > \dfrac{l}{c}$

Using the same reasoning previously detailed in Section 3.2, after the detection of the first photon on the left side of the interferometer, the second photon is located on the right side of the interferometer. However, since the delay is bigger than the time to travel from beam splitters $C$ or $D$ to the detectors $D_j$, it follows that at the time of the detection of the first photon, the second photon has yet not passed the beam splitter $D$. Furthermore, due to the collapse axiom of the wavefunction, after the detection of the first photon, the second photon must be *either* in arm $a_2$ *or* in arm $b_2$; that is, the second photon *is not* in superposition in those arms. Specifically, we know that the photon has a fifty percent chance to be in arm $a_2$ and a fifty percent chance to be in arm $b_2$. Therefore, the photon is in equal mixed states in arms $a_2$ and $b_2$. Then, the photon state is given by the density state

$$\hat{\rho}_M = \frac{1}{2}\left(|1\rangle_{a_2 a_2}\langle 1| + |1\rangle_{b_2 b_2}\langle 1|\right) \qquad (20)$$

$$= \frac{1}{2}\left(\hat{a}^\dagger_{a_2}|0\rangle\langle 0|\hat{a}_{a_2} + \hat{a}^\dagger_{b_2}|0\rangle\langle 0|\hat{a}_{b_2}\right)$$

The photon wavefunction behind the beam splitter $D$ is

$$\hat{a}^\dagger_{a_2} \xrightarrow{D} \frac{1}{\sqrt{2}}\left(i\hat{a}^\dagger_{d_1} + \hat{a}^\dagger_{d_2}\right)$$
$$\hat{a}^\dagger_{b_2} \xrightarrow{D} \frac{1}{\sqrt{2}}\left(\hat{a}^\dagger_{d_1} + i\hat{a}^\dagger_{d_2}\right) \qquad (21)$$

Then the density state becomes

$$\hat{\rho}_M \xrightarrow{D} \frac{1}{4}\left((i\hat{a}^\dagger_{d_1} + \hat{a}^\dagger_{d_2})|0\rangle\langle 0|(-i\hat{a}_{d_1} + \hat{a}_{d_2}) + (\hat{a}^\dagger_{d_1} + i\hat{a}^\dagger_{d_2})|0\rangle\langle 0|(\hat{a}_{d_1} - i\hat{a}_{d_2})\right), \qquad (22)$$

which reduces to

$$\hat{\rho}_M = \frac{1}{2}\left(\hat{a}^\dagger_{d_1}|0\rangle\langle 0|\hat{a}_{d_1} + \hat{a}^\dagger_{d_2}|0\rangle\langle 0|\hat{a}_{d_2}\right).$$

Clearly the photon will be detected in only one of the detectors $D_{d_1}$ and $D_{d_2}$[9,10]

$$_{d_1}\langle 1|\hat{\rho}_M|1\rangle_{d_2} = 0 \qquad (23)$$

To find which of the detectors the photon will be detected in, we calculate

$$_{d_1}\langle 1|\hat{\rho}_M|1\rangle_{d_1} = {}_{d_2}\langle 1|\hat{\rho}_M|1\rangle_{d_2} = \frac{1}{2} \qquad (24)$$

That is, there is equally probability to detect the photons in either detector.

Then the correlation between the first photon detected at detector $D_{c_1}$ and the second at detector $D_{d_2}$ changes from what is found in Eq. (17) to

$$\mathrm{corr}_{\tau_{1/2}}(D_{c_1}, D_{d_2}) = 1 \rightarrow \mathrm{corr}_{\tau_3}(D_{c_1}, D_{d_2}) = \frac{1}{2}. \qquad (25)$$

The correlation *reduces from 1 to 1/2*.

Similarly, the correlation between $D_{c_1}$ and $D_{d_1}$ changes from what is found in Eq. (15) to

$$\mathrm{corr}_{\tau_{1/2}}(D_{c_1}, D_{d_2}) = 0 \rightarrow \mathrm{corr}_{\tau_3}(D_{c_1}, D_{d_2}) = \frac{1}{2}. \qquad (26)$$

This correlation *increases from* $0$ *to* $1/2$.

Therefore, there is a fifty percent probability that both detectors read positive detection and fifty percent negative detection. The difference between the correlations (17) and (25), and between the correlations (15) and (26) will next be used to establish a test of the instantaneousness of the collapse.

## 4 A test for instantaneous collapse

In Eq. (26) we see that for time delay $\tau_3$ and post selection measurements of one photon detected at $D_{c_1}$, the second photon will be detected at $D_{d_2}$ half of the time. In contrast, for either time delay $\tau_1$ or $\tau_2$, no photon detection occurs at $D_{d_2}$. There is no *ambiguity about the cause*. This establishes an *unambiguous* detectable *cause-and-effect* relationship between the detection of the first photon at $D_{c_1}$ and the detection of the second photon at $D_{d_2}$. By *unambiguous detectable cause-and-effect* relationship we mean that there is no other way to explain the results of detection of a photon at detector $D_{d_2}$ except for the detection of the first photon at detector $D_{c_1}$ the time delay $\tau_3$, and the collapse.

The next question about collapse that we address is whether the wavefunction collapse depends on finite times, that is, a duration, or if it is instantaneous. From Eqs. (25) and (26), we see that the correlation function between the two detections depends on the time delay $\tau$.

Now $\tau_3$ is in fact also the difference between the ideal time detection of the two photons, one at detector $D_{c_1}$ and the other at detector $D_{d_2}$.

Then, the dependence of the correlation function on $\tau_3$ reads,

$$\operatorname*{corr}_{\tau_3 \to \tau_2}\left(D_{c_1}(\tau), D_{d_1}(\tau)\right) = \begin{cases} 0 & \tau = \tau_2 \\ \dfrac{1}{2} & \tau = \tau_3 \end{cases} \qquad (27)$$

Ideally, to detect the discontinuity in Eq. (27), the experiment should be performed by taking the limit of the time delay $\tau_3 \to \tau_2$. According to Eq. (27), if the collapse is indeed instantaneous, there should be a sharp discontinuity, a change in the correlation function, at the limit. A discontinuity is the signature of the instantaneousness of the collapse; if collapse takes finite time, the shape will be smooth. Thus, discontinuity is a verification test for instantaneousness of collapse of the wavefunction.

Clearly, because all detections take place in the configuration space, it follows that it is a test for the collapse in the configuration space. Such a test seems to be new.

Next, we take into account experimental limitations of the proposed test.

## 5 Applying the test to the detectors with finite time of operation

In our gedanken experiment above, we have made some simplifications. For example, in the mechanism of detection and in the beam splitting, photons were described as points. The impact of dropping those simplifications will make the experimental results statistical, at least to a degree. Let's consider one such example to see how the proposed test behaves. Consider the finite operation time of the detectors. The detection process is a physical process and therefore the detection of a photon takes a finite time $\Delta$. If the detectors show positive readings at $t_0$, then according to the axiom of collapse, the collapse must have occurred at time $\xi$ within a window of width $\Delta$, i.e. $\xi \in [t_0, t_0 - \Delta]$. Now, consider a probability distribution $f(\xi)$. In principle, continuous adjustment of the time delay $\tau_3 = l/c \pm \Delta$ can be used to get information about $f(\xi)$. For example, if the distribution $f(\xi)$ is uniform over $\Delta$ then a time delay of $\tau_3 = l/c - \Delta/4$ will reduce the number of counts in $D_{d_2}$ by a factor of $1/8$, and so on.

Consider some elements of the theory of collapse given in Ref. [6], which hypothesizes that collapse is not instantaneous but rather has a finite duration $\delta t$, starting at time $t^{(h)}$. But without "specific statements about the nonunitary time evolution in the interval $[t^{(h)}, t^{(h)} + \delta t]$." The arguments in the previous section for finding $\Delta$ apply similarly to find $\delta t$. Furthermore, even though in the model in Ref. [6] for the evolution during the time interval $[t^{(h)}, t^{(h)} + \delta t]$ has not been specified (see the above quotation), the test given here can be used to investigate such possible evolution. Consider the measurements of the first photon at the detector $D_{c_1}$, that is, consider the first photon measured in the time interval $[t^{(h)}, t^{(h)} + \delta t]$. Then let us ask how the second photon would behave: would it behave according to the correlation given in Eq. (15), or would it behave according to correlation in Eq. (26), or perhaps nether of these possibilities.

5. Summary

We presented an experimental test for the instantaneousness of collapse of a wavefunction in configuration space.

The steps taken to construct the test are as follows:

- In Fig. 1, the interferometer for two identical photons was presented.
- Then the post-selected measurements of the first photon at detector $D_{c_1}$ is taken (see the final paragraph of Section 2)
- We showed that for time delay $\tau_2$, the second photon collapses into leg $|\ \rangle_{d_2}$, Eq. (19), while for time delay $\tau_3$ the second photons collapses in to the mixture of $|\ \rangle_{a_2}$ and of $|\ \rangle_{b_2}$. [Eqs. (25), (26)]
- This, in turn, causes the second photon to change its behavior, depending on the time delay: it reduces the probability of measuring the photon at detector $D_{d_1}$ and increase the probability to measuring the photon at detector $D_{d_2}$.
- Since the time delay, $\tau_3$, is a continuous parameter, it has been used to examine the collapse time and its instantaneousness. In particular, we showed the connection between a discontinuous change in a correlation function and an unambiguously detectable cause-and-effect measurement. (27)
- Finally, Section 5, a more realistic model of detectors, with fewer simplifying assumptions, was included in the model to show how better to test instantaneousness experimentally.

Since all the measurements are in the special coordinates of the particles, the test is for the collapse of the wavefunction in configuration space. Although in the present paper the test was for two photons, the same experiment may be performed for two massive bosons, as is typically done in the field of atom optics [13]. Note, once again, the major difference between the present test of collapse as compared to the test of collapse in the anti-bunching tests: The anti-bunching tests are based on correlation measurements. Here, in contrast, we introduced a mechanism in which the photons collapse into a new configuration state, a mixture whose outcome can be measured, which gives a test of the instantaneous collapse time. Note that our test leans on post selected measurements, that is, on knowing that the first photon has been detected at $D_{c_1}$. Therefore no claim of superluminous transfer of information is relevant. The relation of unambiguously detectable cause-and-effect measurement to the Bell inequality needs further study.